\documentclass[a4paper]{jpconf}
\usepackage{graphicx}
\begin{document}
\title{First Results of the Full-Scale OSQAR Photon Regeneration Experiment}

\author{M Schott$^1$, P Pugnat$^2$, R Ballou$^3$, L Duvillaret$^4$, T Husek$^5$, R Jost$^6$, L Flekova$^7$, M Finger Jr.$^5$, M Finger$^5$,  J Hosek$^7$, M Kral$^7$, K Macuchova$^7$, K A Meissner$^8$, J Morville$^9$, D Romanini$^6$, A Siemko$^1$, M Slunecka$^5$, M Sulc$^{10}$, G Vitrant$^4$, and J Zicha$^7$ (OSQAR collaboration)}

\address{$^1$ CERN, CH-1211 Geneva-23, Switzerland}
\address{$^2$ LNCMI-G, CNRS-UJF-UPS-INSA, BP 166, 38042 Grenoble Cedex-9, France}
\address{$^3$ Institut N\'{e}el, CNRS and UniversitŽ Joseph Fourier, BP 166, 38042 Grenoble Cedex-9, France}
\address{$^4$ IMEP-LAHC, UMR CNRS 5130, Minatec-INPG, 3 parvis Louis N\'{e}el, BP 257, 38016 Grenoble Cedex-1, France}
\address{$^5$ Charles University, Faculty of Mathematics and Physics, Prague, Czech Republic}
\address{$^6$ LSP, UMR CNRS 5588, UniversitŽ Joseph Fourier, BP 87, 38402 Saint-Martin d'H\`{e}res, France}
\address{$^7$ Czech Technical University, Faculty of Mechanical Engineering, Prague, Czech Republic}
\address{$^8$ Institute of Theoretical Physics, University of Warsaw, Poland}
\address{$^9$ LASIM, UMR CNRS 5579, Universit\'{e} Claude Bernard Lyon-1, 69622 Villeurbanne, France}
\address{$^{10}$ Technical University of Liberec, Czech Republic}

\ead{mschott@cern.ch}

\begin{abstract}
Recent intensive theoretical and experimental studies shed light on possible new physics beyond the standard model of particle physics, which can be probed with sub-eV energy experiments. In the second run of the OSQAR photon regeneration experiment, which looks for the conversion of photon to axion (or Axion-Like Particle), two spare superconducting dipole magnets of the Large Hadron Collider (LHC) have been used. In this paper we report on first results obtained from a light beam propagating in vacuum within the 9 T field of two LHC dipole magnets. No excess of events above the background was detected and the two-photon couplings of possible new scalar and pseudo-scalar particles can be constrained to be less than $1.15\cdot10^{-7} \mbox{GeV}^{-1}$ and $1.33\cdot10^{-7} \mbox{GeV}^{-1}$ respectively, in the limit of massless particles.
\end{abstract}

\section{Introduction}
\label{intro}

The axion is a neutral pseudo-scalar particle predicted independently by S. Weinberg \cite{T1} and 
F. Wilczek \cite{T2} from the Peccei and Quinn symmetry breaking \cite{T3}. It remains the most plausible solution to the strong-CP problem and constitutes a fundamental underlying feature of the string theory in which a great number of axions or Axion-Like Particles (ALPs) is naturally present \cite{T4}. In addition, the interest in axion search lies beyond particle physics since such hypothetical light spin-zero particles are considered as one of the most serious dark-matter candidates \cite{T5}, and the only non-supersymmetric one. Within this scope and in agreement with previous measurement results excluding "heavy" axions \cite{T6}, the allowed range for the "invisible" axion mass is nominally $10^{-6} < m_A < 10^{-2} eV$.

Most of the experimental approaches to the search for "invisible" scalar or pseudo-scalar particles, such as the axion, are based on the the concept of their coupling to two photons \cite{T7}. The simplest and most unambiguous purely laboratory experiment to look for axion and ALPs is the so-called "photon regeneration" or "shining light through the wall" experiment \cite{T8}, which is of double oscillation type. A linearly polarized laser light beam propagating in a transverse magnetic field is sent through an optical absorber. When the linear polarization of the light is parallel to the magnetic field, photons of energy $\omega$ can be converted to axions with a maximum of probability $P_{\gamma\leftrightarrow A}$ due to the mixing effect. Such weakly interacting particles can then propagate freely through the absorber before being regenerated in the magnetic field on the other side of the absorber. For scalar particles the maximum probability corresponds to a light polarised perpendicularly to the magnetic field.

If a region of length $L$ is permeated by a transverse magnetic field $B$, the photon-to-axion ($\gamma \rightarrow  A$) conversion probability, as well as the axion-to-photon ($A \rightarrow \gammaγ$) one, are given in vacuum by \cite{T8}:

\begin{equation}
P_{\gamma\leftrightarrow A} = \frac{1}{4\beta_A} (g_{A\gamma\gamma}BL)^2 \left( \frac{2}{qL} sin\frac{qL}{2} \right)^2
\end{equation}
with $\hbar=c=1$. Here $\beta_A$ is the axion or ALP velocity, $g_{A\gamma\gamma}$ the axion/ALPs diphoton coupling constant and $q = \|k_\gamma - k_A\|$ the momentum transfer. The energy $\omega$ is the same for photons and axions, $k_A = (\omega^2-m_A^2)^{1/2}$, $b_A = k_A/\omega$ and $k_\gamma=\omega$. The form factor of the conversion probability has its maximum for $q\cdot L = 0$, which corresponds to the limit $m_A \ll \omega$. Otherwise incoherence effects emerge from the axion-to-photon oscillation, reducing the conversion probability. 


\section{The OSQAR Experiment}

The experimental setup of the photon regeneration experiment using two LHC dipole magnets equipped with an optical barrier at the end of the first magnet is schematised in Figure 1. To operate, the LHC dipole is cooled down to 1.9\,K with superfluid He and provides in two apertures a transverse magnetic field which can reach 9.5\,T at maximum over a magnetic length of 14.3\,m. 

As for all LHC superconducting magnets, the dipole used for OSQAR was thoroughly tested at $1.9\,K$. In particular, the field strength and field errors were precisely characterized. An ionized $Ar^{+}$ laser, able to deliver in multi-line mode up to 7 W of optical power is used as light source. The optical beam is linearly polarized with a vertical orientation. To align the polarization of the light in the horizontal direction, a $\lambda/2$ wave-plate is inserted between the laser and the first LHC dipole. It introduces an optical power loss of 20\% at the laser wavelengths. The laser was operated in multi-line mode with approximately 6.4W of the optical power at 514\,nm (2.41\,eV). The laser beam profile was measured at the location of the photon detector and can be well fitted with a gaussian distribution. For photon counting, a $LN_2$ cooled CCD detector from Princeton Instrument is used. It is composed of an array of 1100 pixels of $5\,mm$ height densely packed over a length of 27\,mm. The effective sensitive fraction area of the CCD was increase from 65\% to 100\% by using an optical lens with a focal length of 100\,mm. The quantum efficiency of the detector is equal to $50\pm2\%$ for the Ar+ laser wavelengths and the number of dark counts per pixel is lower than 0.1\,cts/mn. 

\begin{figure}[h]
\includegraphics[width=25pc]{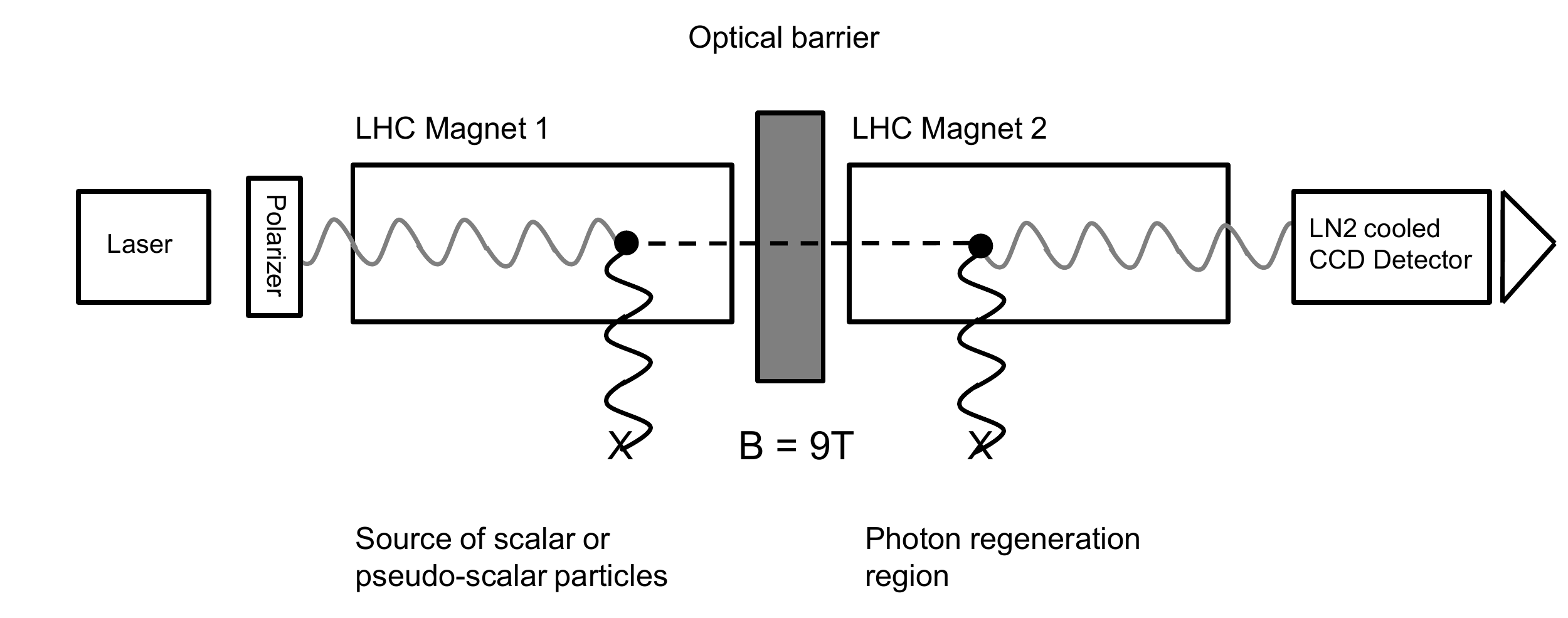}\hspace{4pc}%
\begin{minipage}[b]{8pc}\caption{\label{fig:1}Schematic view of the OSQAR photon regeneration experiment with two LHC magnets.}
\end{minipage}
\end{figure}

\section{Data Taking}

The expected counting rate of photons can be expressed as a function of (1) by:

\begin{equation}
\frac{dN_\gamma}{dt} = \frac{P}{\omega} \eta P_{\gamma \leftrightarrow A}^2
\end{equation}

where $P$ is the optical power, $\omega$ the photon energy and $\eta$ the detector efficiency. It varies as the fourth power of the field integral along the magnet length. The OSQAR photon regeneration experiment was performed with a maximum residual gas pressure lower than $10^{-3}$\,mbar at a temperature around $20^\circ$C. When switched on, the magnetic field was settled to 9\,T. The average optical power of the laser was recorded during measurements and averaged values of 6.4\,W and 6.7\,W were measured for the polarisation parallel and perpendicular to the magnetic field respectively. 

\begin{figure}[h]
\includegraphics[width=26pc]{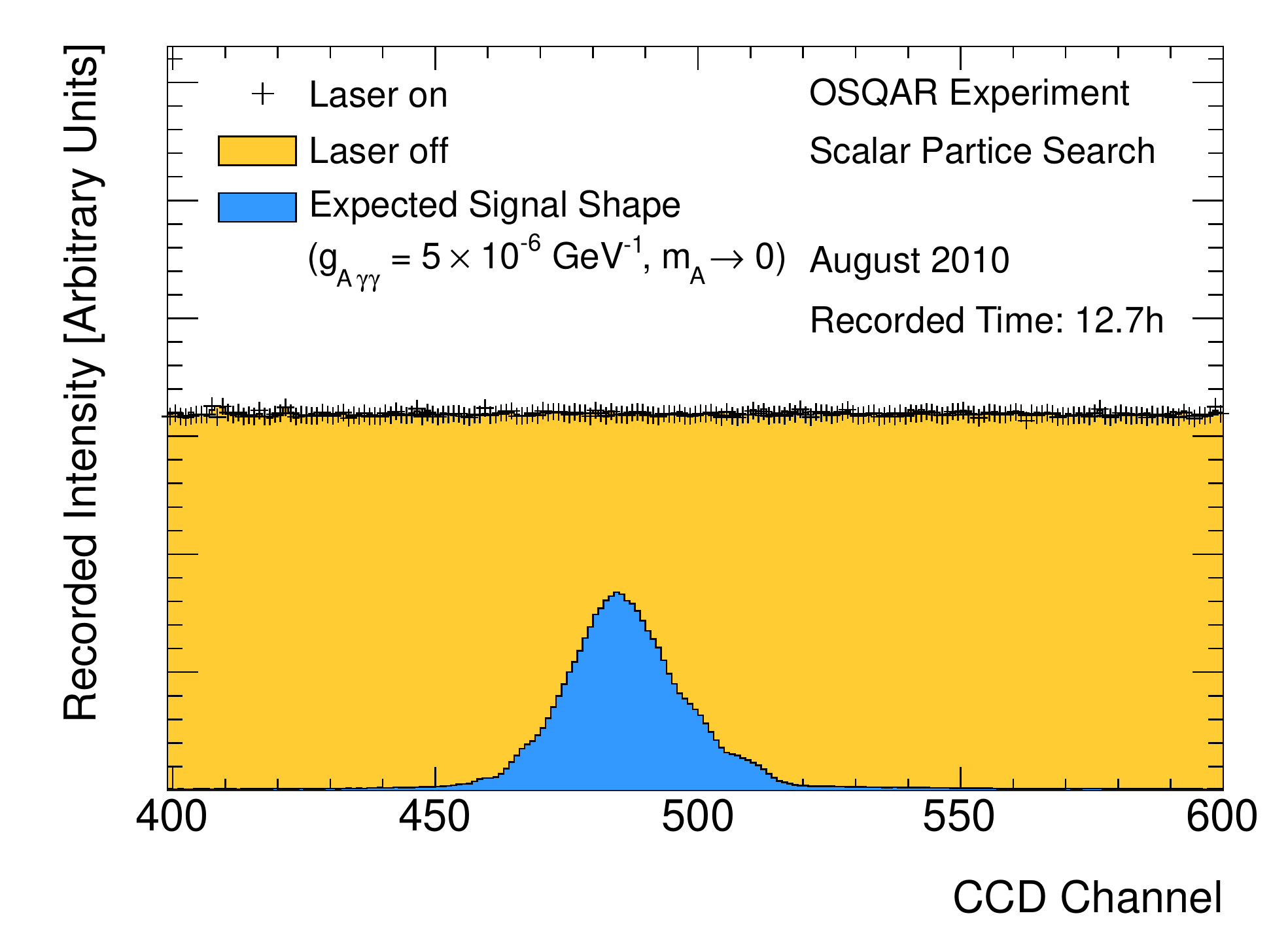}\hspace{3pc}%
\begin{minipage}[b]{9pc}\caption{\label{fig:DataDemo}Recorded spectra of CCD camera for channels 400 to 600, corresponding to an integrated data-taking time of 12.7h, when using $\lambda/2$ wave-plate. The normalized expected background shape for the given integration time is also shown, as well as the expected signal shape, corresponding to $g_{A\gamma\gamma}=5\cdot10^{-6}GeV^{-1}$ in the massless particle limit.}
\end{minipage}
\end{figure}

The integration time for the CCD detector was chosen to be 120\,s to 300\,s for each run of data-taking. These relatively short time interval allowed a clean identification and removal of cosmic rays, which gave rise to a large signal in very few neighboring CCD pixels. The acquisition of photons was performed for various states of the experiment and the runs with turned on and turned off laser have been taken in alternating order. The taken spectra for the runs with laser on/off and with/without the $\lambda/2$ wave-plate have been added up, respectively. The resulting spectra with $\lambda/2$ wave-plate used is shown in Figure \ref{fig:DataDemo}, corresponding to 12.7h integrated data-taking time. The same plots shows the expected background shape, i.e. data taken with the laser turned off, which was normalized to match the given integration time. The expected shape of a regenerated photon signal was also added schematically in Figure \ref{fig:DataDemo}. It is assumed that the signal shape of regenerated photons is close to the recorded laser spectrum when removing the barrier between the two LHC magnets. The overall shape can be described by a gaussian function to a good approximation; the expected signal region was to a range which contains more than 90\% of the expected regenerated photons.

The total integration time of data-taking with optical beam on, magnetic field $B$ on and the polarization of the light was 23.8h. The integration time with optical beam on, magnetic field $B$ on and without the polarization of the light was 5.5h, while the integration time of the background like sample was 20.9h. 

\section{Data Analysis and Exclusion Limits}

The state without optical power nor magnetic field allows characterizing the true background signal. The uniform background noise coming mostly from the CCD readout noise and dark currents gives an integrated value of $\approx42\pm1$ entries per pixel and second. The recorded data with the optical beam, the magnetic field $B$, and the polarization of the light aligned either parallel or perpendicular to $B$ allow probing the existence of pseudo-scalar and scalar particles respectively once the background signal is subtracted. The background in the signal region of the CCD can be estimated in two ways: One approach is based on the recorded background spectra (magnetic field and laser turned off) which is normalized to the analyses signal sample, without taking the signal region into account. The expected background rate can then be estimated via the normalized background spectra in the signal region. Another approach is exclusively based on data, which was taken with turned-on laser, by extrapolating the expected background from the sidebands of the recorded spectra into the signal region. Both approaches have been applied and lead to consistent results. No significant excess of events above the background can be detected. 

\begin{figure}[h]
\begin{minipage}{18pc}
\includegraphics[width=18pc]{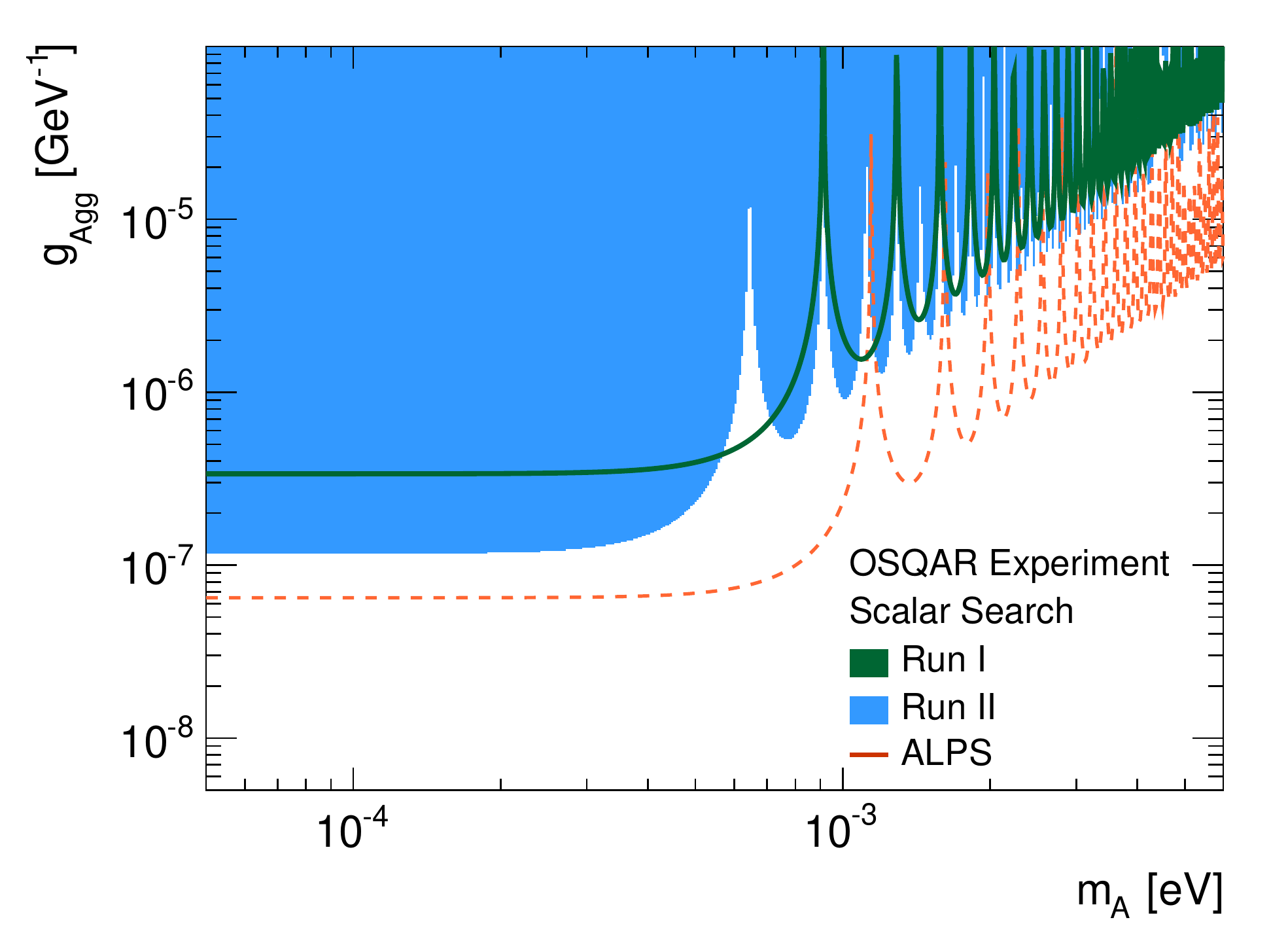}
\caption{\label{Fig:S}OSQAR exclusion limits for scalar ALPs in vacuum for the 2010 run using two LHC dipole magnet. The latest results of the ALPs experiment as well as the previous OSQAR results are also indicated.}
\end{minipage}\hspace{2pc}%
\begin{minipage}{18pc}
\includegraphics[width=18pc]{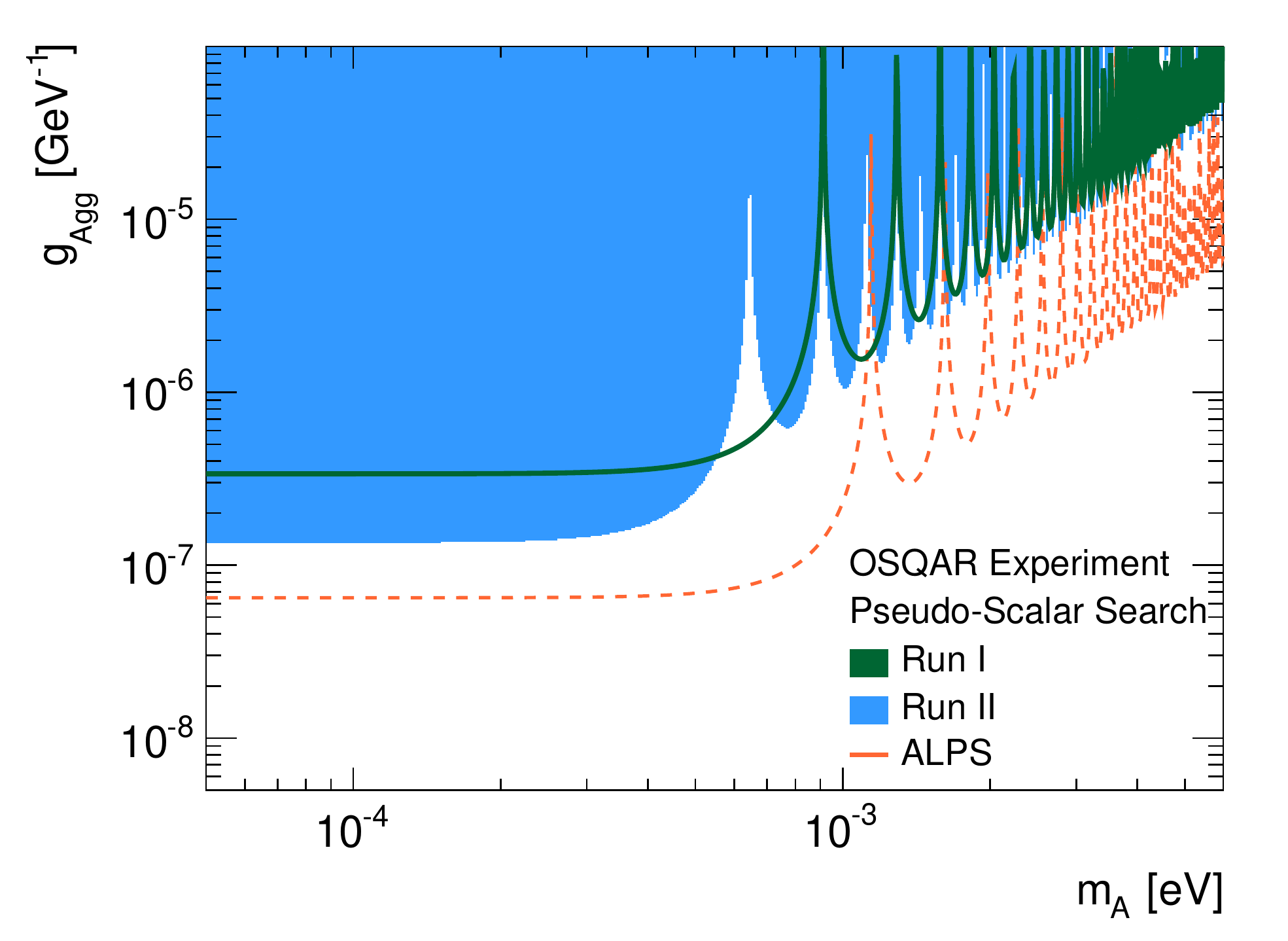}
\caption{\label{Fig:PS}OSQAR exclusion limits for pseudo-scalar ALPs in vacuum for the 2010 run using two LHC dipole magnet. The latest results of the ALPs experiment as well as the previous OSQAR results are also indicated.}
\end{minipage} 
\end{figure}

The exclusion region at 95\% confidence level for the OSQAR run II in vacuum is given in Figure 2. It corresponds to purely laboratory experiment searches of scalar and pseudo-scalar particles that can couple to photons and has been calculated numerically by solving the Equation 2 with a photon flux detection threshold at 95\% of confidence level equal to 0.013 photon/s for the scalar particle search and 0.033 photons/s for the pseudo-scalar particle search, respectively. The difference in the given sensitivities rises from the difference in the integrated data taking time and also from a non-significant upwards fluctuation in the pseudo-scalar particle search. In the limit $m_ A\ll\omega$ the constraints obtained on di-photon coupling constants are $g_{A\gamma\gamma}< 1.15 \cdot10^{-7} GeV^{-1}$ and $g_{A\gamma\gamma}<1.33\cdot10^{-7} GeV^{-1}$ for scalar and pseudo-scalar particles respectively. The 95\% of confidence level exclusion limits for varying axion masses $m_A$ are illustrated in Figure \ref{Fig:S} and \ref{Fig:PS} for scalar and pseudo-scalar particles, respectively.

\section{Conclusion}

The results of the OSQAR photon regeneration experiment obtained within vacuum have been presented. These results are allowing the test of PVLAS results \cite{T11}, \cite{T12} and are consistent with those published by BFRT \cite{T13}, BMV \cite{T14}, LIPPS \cite{T15},  GammeV \cite{T17} and ALPS \cite{T16}. The OSQAR experiment was the first one to confirm the latest and present reference results by ALPS \cite{T16} up to constraints $g_{A\gamma\gamma}< 1.15\cdot10^{-7}\,\mbox{GeV}^{-1}$ and $g_{A\gamma\gamma}<1.33\cdot10^{-7}\,\mbox{GeV}^{-1}$ for scalar and pseudo-scalar particles respectively in the limit of massless particles.

\end{document}